\documentclass[]{appolb}
\usepackage{graphicx}
\usepackage[utf8]{inputenc}
\usepackage[toc,page]{appendix}
\usepackage{amsmath,amsfonts,amssymb}
\usepackage{bm} 
\usepackage{calc}  
\usepackage{color}
\usepackage{enumerate} 
\usepackage{epigraph} 
\usepackage{epstopdf} 
\usepackage[english]{babel}
\usepackage[numbers,sort&compress]{natbib} 
\usepackage{ntheorem} 

\usepackage{pstricks}
\usepackage[percent]{overpic}  
\usepackage{slashed} 
\usepackage{subcaption} 
\usepackage{mymacros} 
\usepackage{cleveref}

\begin{document}
\title{Vector Boson Scattering studies in CMS: \\ The $pp \to ZZ jj $ 
channel\thanks{Presented at the ``Final HiggsTools Meeting'' on September 12$^{th}$ 2017 at the Institute for Particle Physics Phenomenology (IPPP), Durham} 
}
\author{Raquel Gomez-Ambrosio\thanks{raquel.gomez-ambrosio@durham.ac.uk}
\address{\scriptsize{Institute for Particle Physics Phenomenology, Durham University, Durham DH1 3LE, UK, Università di Torino and INFN sezione di Torino. Via Pietro Giuria 1, 10125.Torino, Italy.  }}}

\maketitle
\begin{abstract}
In these proceedings we discuss the family of Vector Boson Scattering (VBS) processes, in particular we look at a very recent result from the CMS collaboration. In this analysis, published in ref.\cite{Sirunyan:2017fvv},  
a search was performed for VBS in the four-lepton and two-jet final state using proton-proton collisions at 13 TeV. The electroweak production of two Z bosons in association with two jets was measured with an observed (expected) significance of 2.7 (1.6) standard deviations, using a multivariate  classifier. Additionally an  expected significance of 1.2 standard deviations was found using matrix elements techniques. Here we will discuss the latter approach in detail. 

\end{abstract}
  
\section{Introduction}

The class of VBS processes refers to the t-channel exchange of two weak bosons  between two quarks, or a quark and an antiquark. Such processes represent a very interesting scenario to measure triple and quartic gauge couplings. Our knowledge of the electroweak sector is rather limited, since the experimentally measured values for these couplings are only constrained to about a 20\% precision.  Studying this group of couplings might shed light on the question of why the EWSB scale is what it is ($v \approx 250 \GeV$) or why the fermions have the experimentally measured masses.

\section{VBS and Unitarity}

\begin{figure}
\begin{center}
\includegraphics[scale=0.2]{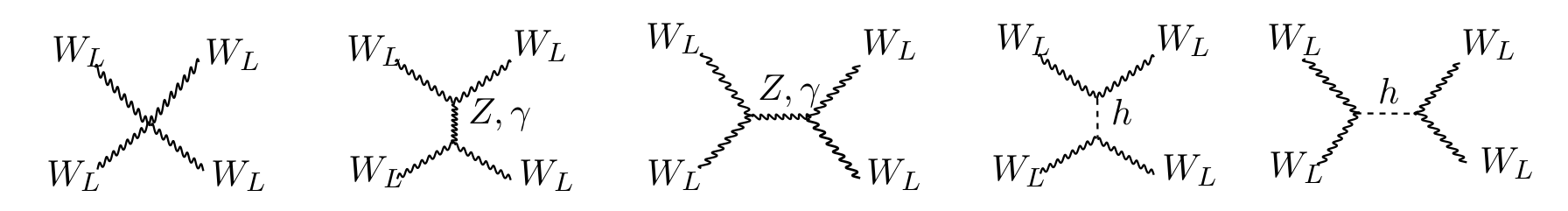}
\caption{Feynman diagrams for the scattering of two longitudinally polarized W bosons of oppostie charge.} \label{fig:WLWL}
\end{center}
\end{figure}

The paradigmatic example to look at within the VBS is the scattering of longitudinally polarized $W^+ W^-$ bosons. In this channel, we can see very easily that the S-Matrix is not unitary\footnote{This behaviour is not unique for this process, it happens in other EW channels with Higgs bosons in the intermediate states.} at high energies until the Higgs boson is included. 
Summing all the purely gauge diagrams we find an amplitude proportional to the centre of mass energy, which makes it diverge at high energies. This behaviour is only cured by the inclusion of a Higgs boson in the t and s channels (last two diagrams of \cref{fig:WLWL}).  
This behaviour is predicted by the Low Energy Theorem (LET) of refs.\cite{Weinberg:1966kf,Chanowitz:1986hu,Chanowitz:1987vj} that describes pion scattering at high energies, the reason why VBS and pion scattering are related lays on the Goldstone boson equivalence theorem of refs.\cite{Cornwall:1974km,Vayonakis:1976vz}.

An interesting phenomenon that should be carefully studied in the context of VBS, unitarity and EFT is that of \emph{delayed unitarity}, proposed in ref.\cite{Ahn:1988fx}. The proposal is to study the behaviour of the scattering amplitude in other high energy regimes, above the one accessible to us but below the artificial $s \rightarrow \infty$. An extension of the SM Lagrangian with a gauge-invariant heavy sector could make such amplitudes (concretely the one for $e^+ e^- \rightarrow W^+_L W^-_L$) to grow again, especially through the radiative corrections to the gauge boson vertices.  If this is the case, we can say that unitarity is \emph{delayed}, predicting an enhancement of the total cross-section for the process that could be measured in experiment as a hint for new physics. 

In order to find concrete unitarity bounds for a process it is useful to do a partial wave expansion of its amplitude, as it was done in refs.\cite{Jacob:1959at,Passarino:1990hk}, starting from the unitarity of the S-matrix,

\begin{align} \label{eq:unitarity1}
& S^{\dagger} S = ( \mathbb{I} - i {\rm T}^{\dagger	}) ( \mathbb{I} + i {\rm T}) =
\mathbb{I} + {\rm T}^{\dagger} {\rm T} + i ({\rm T} - {\rm T}^{\dagger}) 
\nonumber \\ &
 S^{\dagger} S = \mathbb{I} \quad \Rightarrow \quad   {\rm T}^{\dagger} {\rm T} = -i ({\rm T} - {\rm T}^{\dagger})
\end{align}
one can do a partial wave expansion of T, here for $2 \rightarrow 2$ scattering, 

\begin{equation}\label{eq:unitarity2}
\langle f | T | i \rangle = 
16 \pi \sum_{J} (2 J + 1) e^{i (\lambda_{12} - \lambda_{34} ) \phi} d_{\lambda_{12} \lambda_{34}}^J 
\langle f | T^J (E) | i \rangle  
\end{equation}
where $\lambda_{12} = \lambda_1 - \lambda_2$, $\lambda_{34} = \lambda_3 - \lambda_4$ are the initial and final polarizations, $d_{ij}^J (\theta) $  the \emph{Wigner d-functions}. Therefore,  
$a_{\lambda_{12} \lambda{34}^J} =  e^{i (\lambda_{12} - \lambda_{34} ) \phi} \langle f | T^J (E) | i \rangle$ is the J$^{th}$ partial amplitude. Using the completeness relation for the Wigner functions we can find the unitarity condition for each partial wave of the $2 \rightarrow 2 $ amplitude to be,

\begin{equation}
\vert \mathcal{R} ( a_{\lambda \kappa}^{J}) \vert \leq \frac{1}{2} 
\end{equation}
Using this unitarity condition one can put bounds on anomalous couplings, as it was done in refs.\cite{Baur:1987mt,Gounaris:1993fh}. This approach has been extended to partially accommodate EFT (i.e. considering the effect of EFT operators on the gauge couplings but not on the full SM Lagrangian) in refs.\cite{Degrande:2012wf,Corbett:2014ora}. From this point of view, it would be particularly interesting to do the same kind of studies, including all possible contributions from the \dimsix \, EFT basis in the VBS scattering amplitudes. 

\section{Effective Field Theory in VBS}

The study of anomalous triple and quartic gauge couplings (aTGCs and aQGCs) has been a topic of interest for the LHC community since the beginning of its experimental programme independently of the developments in the EFT field. The searches for \emph{forbidden} couplings like $ZZ \gamma \gamma$, $Z \gamma \gamma \gamma$ and $Z Z Z \gamma$ represent an interesting portal for physics beyond the SM and for the study of the EWSB mechanism. However, in the case of VBS where unitarity is preserved thanks to a very precise collection of cancellations between divergent terms, a small change in any of those terms will spoil that equilibrium, and hence this kind of \emph{ad-hoc} variation of the couplings is not the most rigorous approach. 

Some attempts have been done within the EFT community to associate those anomalous couplings to concrete operators in the context of EFT, for example in refs.\cite{Eboli:2006wa,Eboli:2016kko}, and experiments have published different bounds on the values of such operators, for example, in ref.\cite{Sirunyan:2017fvv}. In general the triple gauge couplings are parametrized in terms of \dimsix \, operators, while the quartic ones are commonly written in terms of \dimeight \, contributions. The same way, the experimental collaborations study the quartic gauge couplings in the VBS process and the triple gauge couplings in the rest of the multiboson processes.

This is usually done because the same \dimsix operators contribute to both triple and quartic gauge couplings, while \dimeight contribute only to the QGC, however it is obvious that it is non consistently in therms of the field theory to \emph{skip} a perturbative order in favour of the following one. 

\section{Experimental searches for VBS: State of the art}

Experimentally the family of VBS processes has very particular signatures: there are two forward jets, which are very energetic and have no hadronic activity with each other (no gluon exchange), and in the central region of the detector lay the decay products of the VBS interaction. 
There is also a characteristic rapidity gap between the vector bosons. 
These are the main features that allow to discriminate the VBS signal in the experiment (by \emph{tagging} the two jets). These features play also a fundamental role in Higgs studies, to isolate VBF from other production modes.

\par 
Still, the VBS cross-section is around three orders of magnitude smaller than those for the other common processes at LHC (ie. $\op(\rm{fb})$ instead of $\op(\rm{pb})$), the expected cross-section improves significantly when going from $8 \TeV$ to $13 \TeV$.  For example, for the $pp \rightarrow ZZjj \rightarrow 4 \ell jj$ case, the LO integrated cross-sections for the standard experimental fiducial volumes are, 

\begin{equation}
 \sigma_{LO} (pp \rightarrow ZZjj \rightarrow 4 \ell jj) =  \underbrace{59.79 \pm 0.05 \, ab}_{\sqrt{s} = 8 \TeV} 
 \quad \underbrace{= 228.90 \pm 0.16 \, ab}_{\sqrt{s} = 13 \TeV}  
\end{equation}
the NLO-QCD results for most VBS channels are already available, and predict moderate K-factors in the kinematic regions where LHC-searches look for this process. In this particular case it has been calculated to be $K= 1.02$, in refs.\cite{Denner:2012dz,Figy:2003nv}.

The main difficulties when searching for these processes at LHC are large QCD backgrounds. Electroweak backgrounds, can generally be controlled with experimental cuts. For the $VBS(ZZ)$ case, the QCD-induced background has a much larger cross-section than the actual VBS production, as measured in ref.\cite{Sirunyan:2018vkx}. However there are some privileged channels, like the one with same-sign $W$ bosons in the final state, $VBS(ssWW)$, where the signal-to-background ratio is around 1. This is the only VBS channel that has been observed so far at the LHC\footnote{Update: At the time of the presentation of these contribution in 2017. A handful of other channels was observed during 2018. Updated results can be found in \cref{table:VBS}}. It has been observed at CMS with the $\sqrt{s} = 13 \TeV$ dataset. Additionally there is evidence for the  same observation in ATLAS, as well as for the $Z \gamma$ final state in CMS. The values of the current observed and expected significances can be found in \cref{table:VBS}.

\begin{table}[h!]
\begin{center}
\begin{tabular}{| c | c |  c  |  c |}
\hline
Process	 &  Studied  & Observed 	& Expected  \\
		 &	at $\sqrt{s}$		  & Significance & Significance \\
 \hline 
$Z \gamma$ ATLAS  &  8 TeV  \& 13 TeV    & 2.0  						&  1.8 		\\
$Z \gamma$ CMS  &  8 TeV      &   3.0   						&  2.1		\\
\hline
$W^{\pm}  W^{\pm }$ ATLAS  &  8 \& 13 TeV     & 3.6 \& 6.9				&  2.3		\& 4.6 \\
$W^{\pm}  W^{\pm }$ CMS  &  8 \&  13 TeV      &   2.0 \& 5.5 		&  3.1 \& 5.7		\\ 
\hline
$W^{\pm}  \gamma $ CMS  &  8 TeV      & 2.7				&  1.5		\\
\hline
$Z Z $ CMS  &    13 TeV      &   2.7 		&  1.6		\\ 
\hline
$W^{\pm} Z $ ATLAS  &   8 \& 13 TeV      &  - \& 5.6 		&  - \& -		\\ 
$W^{\pm} Z $ CMS  &   8 \& 13 TeV      &   - \& 2.9 		&  - \& 2.7		\\ 
\hline
\end{tabular}
\caption{Searches for VBS at LHC, state of the art (July 2018).} \label{table:VBS}
\end{center}
\end{table}

\section{Vector Boson analysis within CMS}

The definition of signals in experimental particle physics is a controversial question, and in VBS the situation is particularly delicate. In terms of Feynman diagrams the definition of a concrete process is clear: all the possible diagrams with the same initial and final state particles have to be taken into account when calculating a cross-section. When trying to define a process in LHC, the situation is not so easy and it is necessary to define the concept of ``signal'' where some set of diagrams of interest is isolated from the rest (the ``background''). Still, such background processes interfere with the pure VBS diagrams, and this interference has to be considered.  

\begin{figure} 
\centering
\includegraphics[scale=0.2]{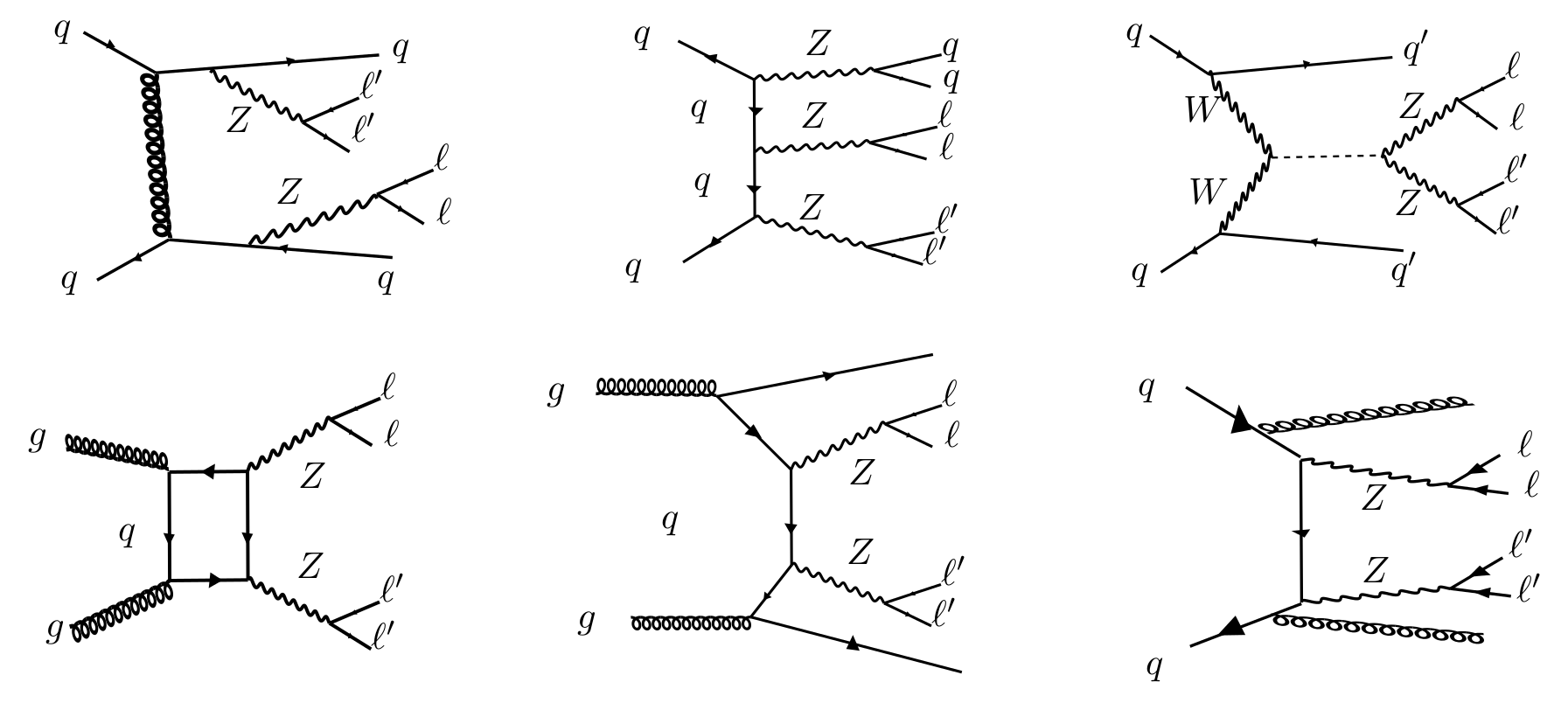}
\caption{On the top row, three VBS-like diagrams that are actually excluded from the definition of the signal. From left to right: QCD induced ZZ production in association with two jets, triboson production, and Higgs-VBF  production.  On the bottom row, the main irreducible backgrounds for the VBS-ZZ channel studied here. From left to right: QCD induced ZZ production (loop and tree contributions) and ZZ production in association with two gluons.} \label{fig:interference}
\end{figure}

\begin{figure}
\centering
\includegraphics[scale=0.2]{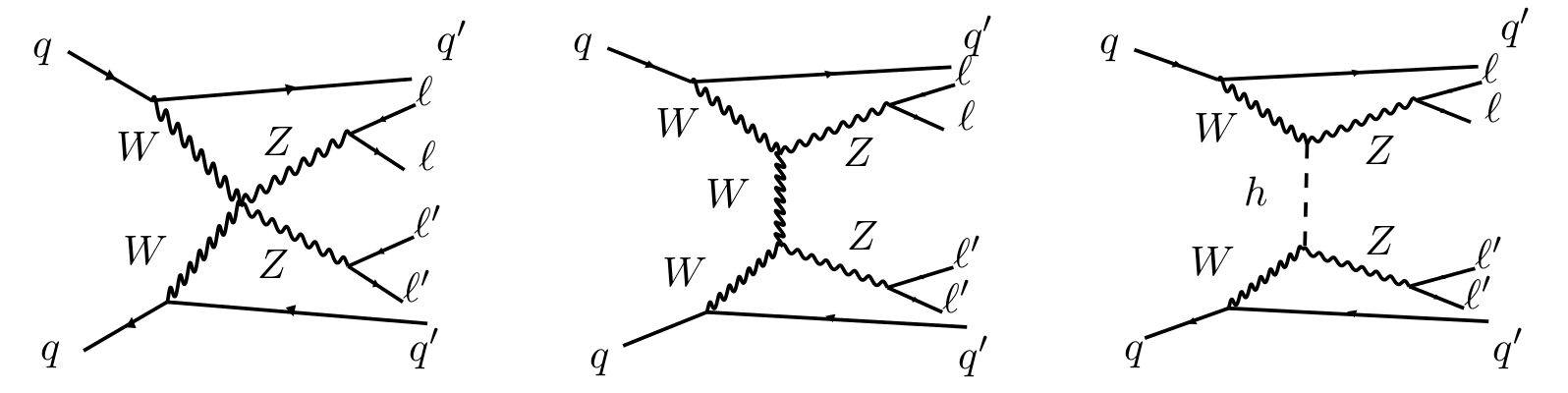}
\caption[Some Feynman diagrams for VBS (ZZ)]{Some representatives of the Feynman diagrams included in the definition of the VBS(ZZ) signal. 
} \label{fig:CMSVBS}
\end{figure}

One of the most important reasons to study VBS in experiment is to obtain information on gauge couplings, for this reason, when the signal is defined it is reasonable to remove the previously discussed contributions: triboson and QCD induced diagrams where the vector bosons only couple to quarks and not among themselves. Additionally, in LHC studies it is customary to remove the Higgs-VBF \emph{contamination} too, since there are dedicated analysis for this channel within the Higgs programme. Some examples of typical signal and background diagrams are shown in \cref{fig:interference,fig:CMSVBS}.

In the analysis, the fiducial region is defined by a standard CMS set of cuts. Additionally a new volume is defined that is expected to be more appropriate for the VBS signal discrimination. In particular, we select leptons coming from the decay of on-shell Z bosons, to remove the Higgs signal: $M_{\ell \ell} \in [ 60 , 120 ] \GeV$. We select the number of jets in the final state to be $ n_{jet} \geq 2$, and the invariant mass of the two leading jets to be $ m_{jj} > 100 \GeV$, to remove the ``triboson'' production.


\section{Interesting VBF/VBS variables:} \label{sec:confronting}

\begin{figure}[]
\begin{center}
\includegraphics[scale=0.5]{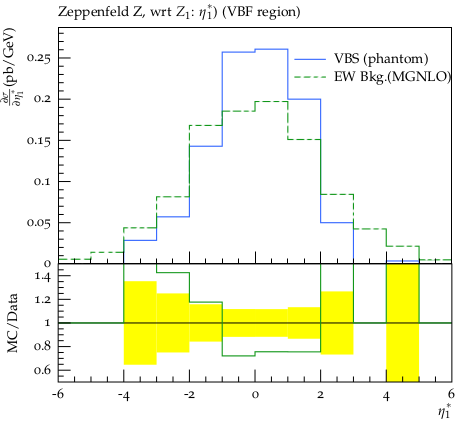}
\includegraphics[scale=0.5]{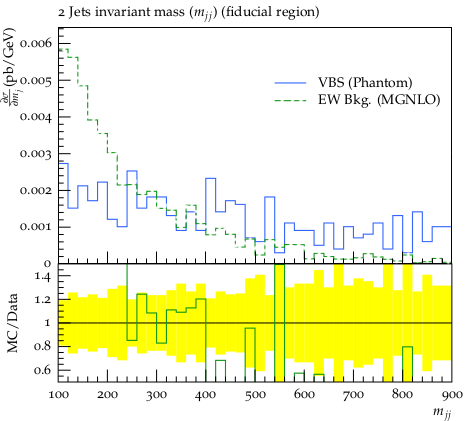}
\includegraphics[scale=0.5]{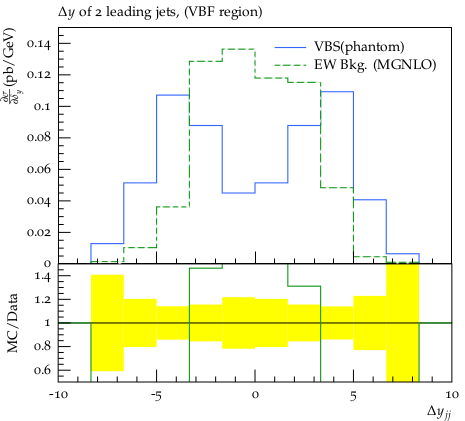}
\caption{Comparison between the VBS(ZZ) signal and the EW-induced background. On the left, $\eta_1^*$ as defined in eq.\eqref{eq:newZepp}. We see that signal and the EW background have a very similar shape in this variable (indeed, it is expected to be optimal at discriminating the QCD background, not the EW one). On the centre we see the dijet mass distributions, which in this case have very particular shapes even in the fiducial region, before the VBF cuts. On the right, Jets $\Delta y$ for the signal and background.  We observe the characteristic ``rapidity gap'' of the signal, and the complementary shape for the background.} \label{fig:zepp12}
\end{center}
\end{figure}

Here we present some variables that are particularly interesting in the VBS analysis. These are: the rapidity difference between the two leading jets ($\Delta y$), the Zeppenfeld variables and the jets mass distributions. We will focus on the shape comparison between signal 
and background (normalizing them to their integrals). The different histograms can be found in \cref{fig:zepp12}. 

\subsubsection*{The Zeppenfeld variable} 
This variable was studied in refs.\cite{Rainwater:1996ud,Govoni:2010bb}, in the context of VBF as a way to isolate the minijets (gluon emission) appearing between the tagged jets. This variable is expected to have very different shapes for the VBF signal and QCD background, the original definition is,

\begin{equation}
\eta^* = \eta_{j_3} - \langle \eta_{j_1 j_2} \rangle , \qquad \text{Zeppenfeld variable}
\end{equation} 
Often, this variable is also called centrality and in fact its precise expression varies from one analysis to the other. 
For the case of interest here, VBS with two Z bosons in the final state it is convenient to define yet another version of the Zeppenfeld variable,

\begin{equation} \label{eq:newZepp}
\eta_{1}^* = \eta_{Z_1} -   \frac{\eta_{j_1} + \eta_{j_2}}{2} , \qquad 
\eta_{2}^* = \eta_{Z_2} -   \frac{\eta_{j_1} + \eta_{j_2}}{2}
\end{equation}
where $j_{1,2}$ are the two leading jets, and $\eta_{Z_{1,2}}$ are reconstructed from the decay products of the $Z$ bosons.

\section{The VBS(ZZ) analysis in CMS} \label{sec:VBSanalysis}

The CMS detector has an onion shape, divided in different shells containing the different sub-detectors, as well as the central feature of the CMS apparatus: a superconducting solenoid of 6 m. internal diameter providing a magnetic field of 3.8 T. The analysis discussed here used a data sample recorded by the CMS experiment during 2016, corresponding to an integrated luminosity of $\Lag = 35.9 fb^{-1}$. For details on the specific MC sample choice as well as the event selection, object reconstruction and trigger selections we refer to the official publication.

%

\subsubsection*{Multivariate Analysis techniques: Matrix Elements}

In order to increase the sensitivity of the analysis given the low expected signal yield, the signal is extracted from a one-dimensional template fit to an MVA output spectrum, implemented as a boosted-decision-tree (BDT) the details of which can be found in ref. \cite{Sirunyan:2017fvv}. Additionally, a Matrix Element (ME) method was implemented  for this analysis, which will be discussed here. 

The ME analysis is based on the study of the processes at the generator level (even before parton shower), in contrast with a classical analysis that studies kinematic distributions at the detector level. The main idea in any ME implementation is to define a discriminant based on probabilities for the appearance of signal and background and, in the case where it leads to better discrimination power than the usual ``cut and count'' approach, use it to extract the signal. 
This method was used already in the extraction of the Higgs signal in the $H \rightarrow ZZ \rightarrow 4 \ell$ analysis and can be a very useful tool in the VBS analysis where the signal is very small.

The advantages of ME techniques are twofold: on one hand, they can (and should) be used as a cross check for the BDT studies, in order to avoid the overtraining of the network that may lead to wrong results. Also, ME techniques provide a theoretical insight on the studied processes, which can be very useful for phenomenology studies in the future if finally the EFT framework is implemented in the official LHC Monte Carlo production.

\subsubsection*{Concept of MELA}  For a given event we chose a representative variable, in this case the $4 \ell$ and $2j$ four-momenta, and we construct probabilities P for it to come from a given process (signal or background). These probabilities P are calculated using matrix elements from MC generators or analytical parametrizations. In this study we used the \texttt{MELA 2.0.1} release, from refs.\cite{Gao:2010qx,Bolognesi:2012mm, Anderson:2013afp,Gritsan:2016hjl}. This, includes MCFM background probabilities at LO QCD coming from: QCD + 2 jets production (second diagram in the bottom row of \cref{fig:interference}) and EW production (last diagram on the bottom row of \cref{fig:interference}). And the signal probabilities for VBS(ZZ) production.  The kinematic input is the final state $4 \ell$ and $jj$ four-momenta, and the baseline selection as that in the analysis.  The signal-background kinematic discriminant is defined as:
\begin{equation}
K_D (M_{4\ell 2j}) = \frac{P_{sig}}{ P_{sig} + P_{bkg} }
\end{equation}
where ``sig'' and ``bkg'' are the two processes we want to separate, and the $P$ are assumed to be normalized to 1. For a given $4\ell$ total mass, there are 7 independent variables for which P are aggregated probabilities, taken correlations into account. In this case with additionally 2 jets, there are 6 more variables.  The main obstacle for this method is that of combining the different background probabilities in one. 

\subsubsection*{Results: ROC curves}

\begin{figure}[h!]
\begin{center}
\includegraphics[scale=0.2]{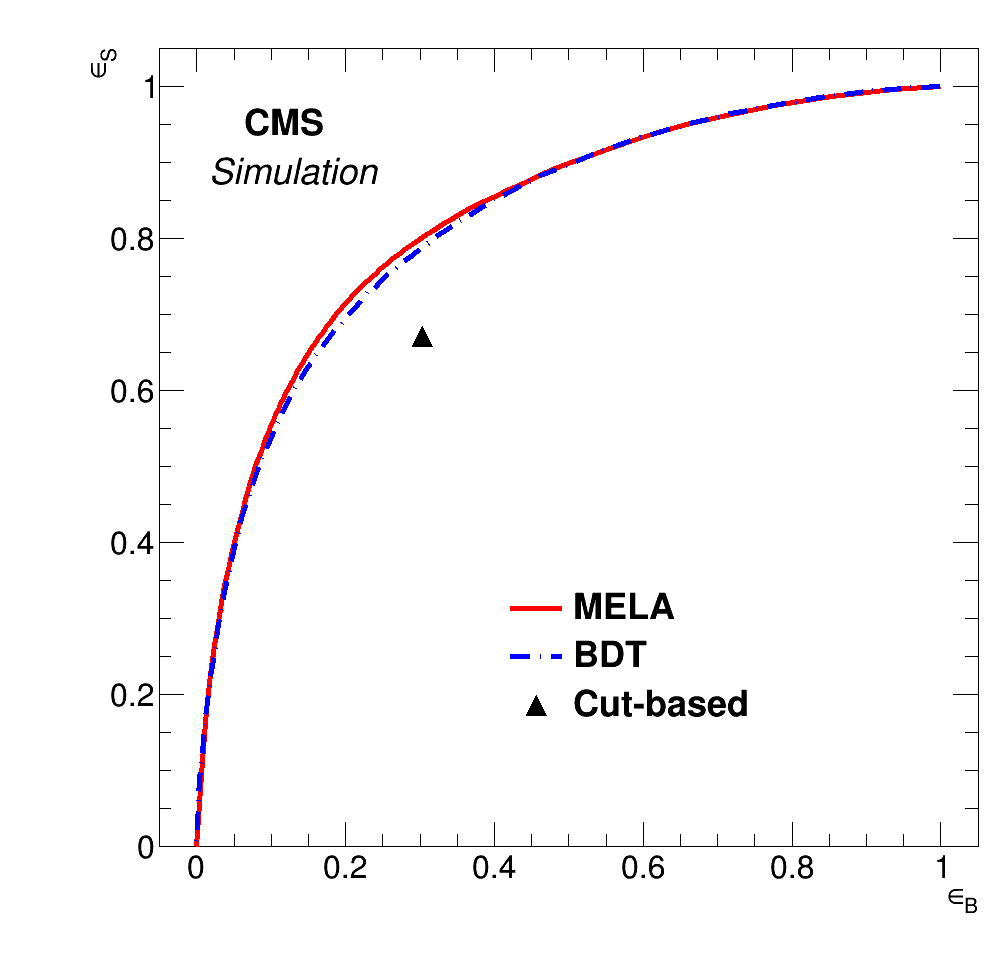}
\end{center}
\caption{ROC curves for the MELA analysis (solid) and the BDT analysis (dashed). The way to interpret this curve is the following: on the Y-axis the signal efficiency  ($\epsilon_S$) is represented, and the X-axis represents the background efficiency ($\epsilon_B$). The points in the ROC curve are the values of $\epsilon_S$ given by the probabilities extracted from the Monte Carlo generator as a function of $\epsilon_B$ (extracted in the same way)  for a scan of different values of the cut on the kinematic discriminant cut (between 0 and 1). The kinematic cuts are represented by a point, which is given as the signal vs. background efficiency, for the best value of the kinematic discriminant ($K_D=0.66$).} \label{fig:ROC}
\end{figure}

The ROC curve (Receiver Operating Characteristic) is a central variable in statistical analysis with binary classifiers. Strictly speaking it is the function of the ``true positive rate'' versus the ``false positive rate'' given by some probabilistic set-up. In particle physics it is used generally for any kind of signal vs. background plots. The ROC curve for this analysis and its description can be seen in \cref{fig:ROC}.

\subsubsection*{Results: Significances and systematic errors}

To extract significances for the observation of the signal the profile likelihood method was used. The implementation we chose of the likelihood fit was the CMS tool \texttt{Combine}.  The theoretical uncertainty is obtained by varying the  renormalization and factorization scales. Uncertainties related with the choice of PDF and strong coupling constant are evaluated following the prescription of the \texttt{PDF4LHC} with the NNPDF sets, refs.\cite{Butterworth:2015oua,Ball:2014uwa}.  The uncertainty on the LHC integrated luminosity for this dataset is $2.6\%$, the trigger efficiency is $98 \% $ for the data and $ 99 \% $ for the MC, hence a systematic uncertainty of $2\% $ is assigned. Uncertainties from lepton reconstruction are $ 6/4/2 \%$ for $4e / 2e2\mu / 4 \mu$ selections. The uncertainty induced by pileup is $4.6\%$,for both signal and background.  The jet energy scale and jet energy resolution were extracted directly from each MC sample for the MELA analysis, and from the MVA template fit in the case of the BDT analysis. They go from $1.12\%$ to $7.24\%$.

The final results for the expected significance of the observation of the VBS(ZZ) signal, is $\sigma_{MELA} = 1.24$. In agreement with the one extracted using the  BDT (1.6 $\sigma$), and given the shape of the ROC curve (\cref{fig:ROC}) it is clear that some improvements on this analysis, mainly a study of the shapes of signal and background, (i.e. a multi-bin fit, based on extracting the maximum likelihood on a bin per bin basis) would lead to the same results as the BDT analysis.  The MELA analysis has the advantage that it is solid from the theoretical point of view (its based on QFT matrix elements) and hence, its output can be analyzed \emph{physically}.

\section{Acknowledgements}

We gratefully acknowledge the input of R. Covarelli, C. Mariotti and G. Passarino to this project. As well as the Research Executive Agency (REA) of the European Union for funding through the Grant Agreement PITN-GA-2012-316704 (“HiggsTools”). The author would also like to acknowledge the contribution of the COST Action CA16108.



\bibliographystyle{atlasnote} 
{\scriptsize
\bibliography{thesisBib}
}


\end{document}